\documentclass[aps,prl,twocolumn,nofootinbib,superscriptaddress,showpacs,floatfix,preprintnumbers]{revtex4-2}

\usepackage[dvipsnames]{xcolor}     
\definecolor{lcolor}{rgb}{0.5,0,0}
\definecolor{citcolor}{rgb}{0,0,1}
\usepackage[breaklinks,colorlinks,urlcolor=blue,citecolor=citcolor,linkcolor=lcolor,linktoc=all]{hyperref}
\usepackage{color}
\usepackage{graphicx}	
\graphicspath{{./figures/}}
\usepackage[utf8]{inputenc}
\usepackage{amsmath} \usepackage{amssymb}
\usepackage[dvipsnames]{xcolor}      
\usepackage[capitalise]{cleveref}
\usepackage{bm,bbm,bbold}

\allowdisplaybreaks

\makeatletter
\g@addto@macro\bfseries{\boldmath}
\makeatother

\usepackage{tikz}
\usepackage[customcolors]{hf-tikz}
\usepackage{mciteplus}

\usetikzlibrary{arrows,cd,shapes,decorations.pathmorphing,decorations.markings,shadings}
\tikzset{
  big arrow/.style={
    decoration={markings,mark=at position 1 with {\arrow[scale=4,#1]{>}}},
    postaction={decorate},
    shorten >=0.4pt},
  big arrow/.default=blue}

\newcommand{\be}{\begin{equation}}
\newcommand{\ee}{\end{equation}}
\newcommand{\bea}{\begin{eqnarray}}
\newcommand{\eea}{\end{eqnarray}}

\newcommand{\mt}[1]{\textrm{\scriptsize #1}}
\def\Nc{N_\mt{c}}
\def\Nf{N_\mt{f}}
\def\Vf{V_\mt{f}}
\def\Vg{V_\mt{g}}
\def\muq{\mu_\mt{q}}
\def\nq{n_\mt{q}}
\def\chiB{\chi^\mt{B}_2}
\def\ie{{\emph{i.e.}}}
\def\eg{{\emph{e.g.}}}

\begin{document}

\title{Is holographic quark-gluon plasma homogeneous?}

\preprint{HIP-2024-12/TH}
\preprint{APCTP Pre2024 - 011}

\author{Tuna Demircik}
\email{tuna.demircik@pwr.edu.pl}
\affiliation{Institute for Theoretical Physics, Wroclaw University of Science and Technology, 50-370 Wroclaw,
Poland}
\author{Niko Jokela}
\email{niko.jokela@helsinki.fi}
\affiliation{Department of Physics and Helsinki Institute of Physics,
P.O.~Box 64, FI-00014 University of Helsinki, Finland}
\author{Matti J\"arvinen}
\email{matti.jarvinen@apctp.org }
\affiliation{Asia Pacific Center for Theoretical Physics, Pohang, 37673, Korea}
\affiliation{Department of Physics, Pohang University of Science and Technology, Pohang, 37673, Korea}
\author{Aleksi Piispa}
\email{aleksi.piispa@helsinki.fi}
\affiliation{Department of Physics and Helsinki Institute of Physics,
P.O.~Box 64, FI-00014 University of Helsinki, Finland}

\begin{abstract}
We present evidence for a spatially modulated instability within the deconfined quark-gluon plasma phase of QCD. This evidence is based on robust predictions from generic holographic models, accurately fitted to lattice data, where the instability is driven by the Chern--Simons term mandated by the flavor anomalies of QCD. Such an instability occurs universally across holographic models 
at surprisingly low densities, within the crossover region amenable to lattice and experimental studies, therefore inviting further explorations of inhomogeneous phases in this region.
\end{abstract}

\maketitle

\section{Introduction}

The beam energy scan (BES) program of the relativistic heavy-ion collision experiment at Brookhaven \cite{Bzdak:2019pkr,Busza:2018rrf,Lovato:2022vgq} is at the forefront of probing one of the most elusive aspects of finite-density Quantum Chromodynamics (QCD): the presence of a critical end point (CEP) in its phase diagram. While transition from the confined hadronic phase to the deconfined quark-gluon plasma (QGP) phase at negligible baryon chemical potentials is characterized as a crossover, the nature of the transition at higher chemical potentials remains unknown. Various models suggest this transition may become a first-order phase transition beyond a certain chemical potential, yet confirming this behavior is challenging due to the sign problem \cite{deForcrand:2009zkb}, which makes this region of the phase diagram inaccessible to lattice simulations. By exploring this critical area, the BES aims to uncover and accurately locate the CEP, shedding light on this pivotal question.

Building on the exploration of the CEP, another intriguing aspect of finite-density QCD involves the investigation of potential ``exotic'' phases within its phase diagram. Among these, phases exhibiting spatial modulation hold particular interest. Theoretical explorations have proposed diverse modulated structures such as the chiral density wave \cite{Deryagin:1992rw,Shuster:1999tn}, crystalline color superconductivity \cite{Alford:2000ze,Mannarelli:2006fy,Rajagopal:2006ig}, chiral magnetic spiral \cite{Basar:2010zd}, and chiral magnetic wave \cite{Kharzeev:2010gd}. These phases are anticipated to reside within the deconfined region of the QCD phase diagram, manifesting at low temperatures and high densities. Additionally, the gauge-gravity duality has suggested another spatially modulated configuration, 
which involves spatially modulated chiral currents
rather than the  chiral condensate~\cite{Buballa:2014tba},
as proposed in the works of Nakamura, Ooguri, and Park \cite{Ooguri:2010xs,Nakamura:2009tf,Ooguri:2010kt} (the NOP phase).

\begin{figure}[h]   
\includegraphics[height=0.28\textwidth]{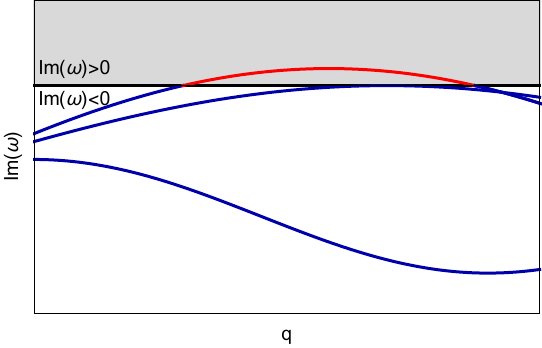}
\caption{\small A schematic plot for the fluctuation mode in the plane wave form $\propto e^{-i(\omega t-qz)}$ giving rise to an instability at nonzero momentum. Blue curves in the lower half plane (Im$(\omega)<0$) represent stable modes and the curve with red segment entering into the upper half plane (Im$(\omega)>0$) denotes an unstable mode. The middle curve arises at the onset of the instability and determines the critical value for the momentum, representing the characteristic length scale of the yet undetermined modulated ground state.}
    \label{fig:p0}
\end{figure}

The presence of this modulated phase is tied to the global flavor anomalies in QCD, which are described by the Wess--Zumino--Witten term in effective theory~\cite{Wess:1971yu,Witten:1983tw}.
In analogy, the global anomalies of QCD dictate 
an essentially unique Chern--Simons (CS) term on the gravity side of the gauge-gravity duality~\cite{Witten:1998qj}. It is this term which drives the NOP phase~\cite{Domokos:2007kt,Nakamura:2009tf,Bergman:2011rf}. The presence of this phase is indicated by an instability of the ``regular'' homogeneous QGP phase, which is modeled through charged hairy black holes in the gauge-gravity duality.  
By studying fluctuations around the black hole geometry, characterized by frequency $\omega$ and momentum $q$, the instability is identified as a fluctuation mode that has positive imaginary part of the frequency at nonzero momentum, as we sketch in Fig.~\ref{fig:p0}.

The NOP instability was originally discussed in the context of QCD~\cite{Ooguri:2010xs} in a specific holographic model, the Witten--Sakai--Sugimoto model, which is based on a certain brane configuration on top of a supergravity background derived from string theory~\cite{Witten:1998zw,Sakai:2004cn,Sakai:2005yt}. 
An alternative approach involves employing generic five-dimensional gravity actions tailored to precisely mirror QCD data. This methodology began with applications to pure Yang--Mills theory and zero-density QCD~\cite{Gursoy:2007cb,Gursoy:2007er,Gubser:2008ny} and was later expanded to include finite densities~\cite{DeWolfe:2010he,DeWolfe:2011ts,Jarvinen:2011qe}.  
Unlike other holographic approaches for QCD,
they accurately reproduce the lattice data for QCD thermodynamics at low densities~\cite{DeWolfe:2010he,Knaute:2017opk,Critelli:2017oub,Jokela:2018ers}, enabling extrapolations to higher densities to study behavior of QCD matter near the postulated CEP \cite{Rougemont:2023gfz} or even at neutron-star matter densities~\cite{Hoyos:2016zke,Jarvinen:2021jbd,Hoyos:2021uff}.

The main purpose of this letter is to analyze the NOP instability in this latter class of generic gauge-gravity duals of QCD. As we shall see, this approach predicts the presence of the instability at surprisingly low densities and high temperatures, near the crossover region of QCD where direct lattice simulations and experimental explorations are feasible. This is interesting because in the low-density region the predictions from this holographic framework are model-independent  to a high degree. 

This letter appears simultaneously with two accompanying articles. In~\cite{QNMpaper} we present a more detailed analysis of the NOP instability in a subclass of models considered here (V-QCD,~\cite{Jarvinen:2011qe}). In~\cite{JJP} we discuss the comparison to lattice data in the other subclasses of models, as we also explain below.

\section{Setup}

We  consider QCD at finite temperature and baryon density with vanishing quark masses and $\theta$-angle. 
We use a generic five-dimensional gravity dual, which includes dual fields for the relevant and marginal operators in QCD. The metric of the gravity theory is dual to the energy-momentum tensor of QCD. The chiral U$(\Nf)_L\times$U$(\Nf)_R$ symmetry of QCD, where $\Nf$ denotes the number of flavors, is lifted to a gauge symmetry on the gravity side. The gauge fields $A_{L}^{\mu\,ij}$ and $A_{R}^{\mu\,ij}$ on the gravity side of the duality, where $i,j =1,\ldots,\Nf$ are the flavor indices, are therefore dual to the left- and right-handed currents in QCD, respectively. Moreover, we introduce a dilaton field $\phi$ which is dual to the $G_{\mu\nu}G^{\mu\nu}$ operator in QCD, where $G$ is the gluon field tensor.

A generic Ansatz for the gravity action can be written as 
\begin{equation}
S=S_\mt{g}+S_\mt{f} + S_\mt{CS} \ .
\end{equation}
The action for the gluon sector is given by five-dimensional dilaton-gravity:\footnote{We use Greek letters $\mu,\nu,\ldots$ (capital Latin letters $M,N,\ldots$) to denote Lorentz indices in four (five) dimensions.}
\begin{equation} 
S_\mt{g}=M_\mt{p}^3\Nc^2\! \int\! \mathrm{d}^5x\,\sqrt{-g}\left[R-\frac{4}{3} g^{M N} \partial_{M} \phi \partial_{N} \phi+\Vg(\phi)\right] \ , 
\end{equation}
where $M_\mt{p}$ is the five-dimensional Planck mass and $\Nc$ denotes the number of colors.
The flavor sector arises from Dirac--Born--Infeld (DBI) action,
\begin{align}
S_\mt{f}&= -\frac{1}{2} M_\mt{p}^3N_c  \! \int\! \mathrm{d}^5x\,\Vf(\phi) &\nonumber\\
&\times\mathrm{Tr}\bigg\{\sqrt{-\operatorname{det}\left[g_{MN}+w(\phi) F_{MN}^{(L)}\right]}+(L \leftrightarrow R)\bigg\}\ , &
\label{sf}
\end{align}
where $F^{(L)} = dA_{L} - i A_{L}\wedge A_{L}$ and $F^{(R)} = dA_{R} - i A_{R}\wedge A_{R}$ are the field strength tensors, and $\mathrm{Tr}$ denotes the trace over the quark flavor indices. The CS term~\cite{Witten:1998qj} is dictated by the flavor anomalies in QCD: 
\begin{align} \label{LCS}
 S_\mt{CS} &= \frac{i\Nc}{24\pi^2} \int \mathrm{Tr}\Big[-iA_L\wedge F^{(L)} \wedge F^{(L)} \nonumber\\
 &\quad\qquad\qquad\qquad+iA_R\wedge F^{(R)} \wedge F^{(R)}+\ldots \Big] \ , 
\end{align}
where we only show the terms relevant for our analysis. 

This action is not the most general two-derivative action that respects the chiral and parity symmetries of QCD because of the choice of the DBI action in~\eqref{sf}. However since it is the gauge fields that control the quark number density $\nq$ 
on the QCD side, their amplitudes are small in the low density region, which is our main interest in this letter. Therefore the flavor action~\eqref{sf} can be well approximated by the leading order truncation of its series expansion at small field strengths $F^{(L/R)}$, \ie, the Yang-Mills action for the gauge fields. Model dependence therefore arises at $\mathcal{O}(F^4)$. We however remind that the DBI form of~\eqref{sf} is the action motivated by string theory.

Finally, the potentials $\Vg$, $\Vf$, and $w$ are determined by comparing to lattice data to the equation of state (EoS) of the pure Yang--Mills theory~\cite{Panero:2009tv}, the EoS for full QCD at zero density~\cite{Borsanyi:2013bia,HotQCD:2014kol}, and the dependence of the EoS on $\nq$~\cite{Borsanyi:2011sw,Bazavov:2020bjn}, respectively. In order to study how much the fitting procedure affects the results, we use different approaches. The first approach is that of~\cite{Gursoy:2007cb,Gursoy:2007er,Jarvinen:2011qe} (improved holographic QCD and V-QCD models), where the gravity dual has a deconfinement phase transition. In this case, we fit data only in the deconfined phase, \ie, above the crossover for full QCD~\cite{Gursoy:2009jd,Jokela:2018ers}. In order to implement this first approach, we use three fits for the V-QCD model (5b, 7a, and 8b)~\cite{Jokela:2018ers,Ishii:2019gta} where the variation between the fits reflects the model dependence that cannot be fixed by the data. In the second approach~\cite{Gubser:2008ny,DeWolfe:2010he} the gravity model is not confining, and one uses the model to fit the data directly at all temperatures, including the data below the crossover. 
In order to implement this second approach, we use the fit from the companion article~\cite{JJP}. We also study a fit where the leading order truncation of the DBI action~\eqref{sf}, \ie, the Yang--Mills action, is directly fitted to data.
These cases cover virtually all available holographic QCD models fitted to lattice data allowing us to assert the universality of the NOP instability in the deconfined QGP phase.

\section{Fluctuating the black hole} 

The background solution at finite density is a five-dimensional charged black hole solution with metric 
\begin{equation} \label{metric}
\mathrm{d} s^{2}
=e^{2A(r)}\left[f(r)^{-1}\mathrm{d}r^{2}-f(r) \mathrm{d} t^{2}+\mathrm{d} \mathbf{x}^{2}\right]\ ,
\end{equation}
and a nonzero temporal component of the vectorial field, $A^t_{L}(r) = A^t_{R}(r) =  \Phi(r)\mathbb{1}$, where $\mathbb{1}$ is the 
unit matrix in flavor space. For black hole solutions, there is a horizon at a finite value of the holographic coordinate $r$ where the blackening factor $f(r)$ vanishes. 
The thermodynamics is then given by the standard black hole thermodynamics, and quark number chemical potential $\muq$ is identified as the value of the $\Phi$-field at the boundary of the geometry where the field theory lives; see the accompanying articles~\cite{QNMpaper,JJP} for more details.

We consider the following infinitesimal plane wave fluctuations on top of the 
finite density background:
\begin{equation}
\begin{aligned}\label{flucs}
\delta A^k_{L}(r,t,z) &=  
e^{-i(\omega t-q z)} \delta A^{ka}_{L}(r)t^a\ , \quad (k = x,y)\ ,
\end{aligned}
\end{equation}
where we chose the momentum to be aligned with the $z$-direction,  and $t^a$ are the $\text{SU}(\Nf)$ generators. 
The fluctuations perpendicular to the momentum ($x$ and $y$ directions) are the only ones receiving contributions from the CS term.
We do not study here the Abelian fluctuations, for which $t^a$ is replaced by the unit matrix in~\eqref{flucs}. Analyzing this sector consistently requires also turning on fluctuations of the metric, and implementing the axial U$(1)$ anomaly in the model~\cite{Arean:2016hcs}. These fluctuations do show the NOP instability, but it turns out to be suppressed with respect to that obtained from the fluctuations we consider here~\cite{QNMpaper}.

The left-handed fluctuations given in (\ref{flucs}) satisfy the equations 
\begin{equation}
\begin{aligned}
\label{eq:LRinstabilityeq}
0=&\,(\delta A_{L}^{\pm\, a})''+\left[\frac{\mathrm{d}}{\mathrm{d} r} \log\left( e^{A}f \Vf(\phi)w(\phi)^2R\right)\right](\delta A_{L}^{\pm\, a})' \\
&+\left[\frac{\omega^2}{f^2}-\frac{q^2}{f R^2}\right]\delta A_{L}^{\pm\, a}  \\& 
\pm q\frac{e^{-2 A} \hat{n}  }{2 \pi^2 M^3_p f   \Vf(\phi)^2w(\phi)^4R^2}\delta A_{L}^{\pm\, a} \ ,
\end{aligned}
\end{equation}
where 
\be 
R=\sqrt{1+\frac{\hat n^2}{e^{6 A} w(\phi)^2 \Vf(\phi)^2}}\ ,
\ee
and $\delta A^{\pm}_{L}=\delta A_{L}^x\pm i \delta A_{L}^y$. 
The constant of integration $\hat n$, which is proportional to $\nq$, arises from the solution of the background temporal gauge field,
\begin{equation} \label{nhatdef}
\hat n=-\frac{e^{A} \Vf(\phi) w(\phi)^{2} \Phi'}{\sqrt{1 -e^{-4A}w(\phi)^2 (\Phi')^2 }} \ .
\end{equation}
The CS term contribution, \ie, the last term in~\eqref{eq:LRinstabilityeq}, gives rise to the instability. Indeed, when both the momentum $q$ and the quark number $\hat n$ are nonzero this term acts as an effective negative mass squared for the fluctuation mode, provided the signs in the equation are appropriately chosen. The right-handed fluctuations satisfy the same equation but with the opposite sign in the CS contribution, indicating that they too exhibit the instability.  

The presence of the instability indicates that the true gravity ground state possesses spatially modulated fields, for example, $A^{xa}_{L}(r,z)$~\cite{Ooguri:2010kt}, which means that the dual QCD phase has modulated currents for quark fields $\psi$. That is, for example,  $\langle\bar \psi\gamma^{x}(1-\gamma_5)t^a\psi\rangle$ depends on $z$.

\section{Results}

\begin{figure}  
\qquad\includegraphics[height=0.042\textwidth]{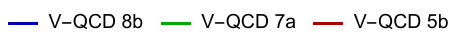}
\\
\includegraphics[height=0.32\textwidth]{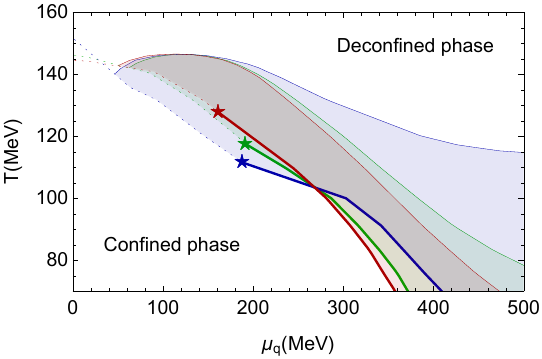}
\includegraphics[height=0.042\textwidth]{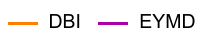}
\\
\includegraphics[height=0.32\textwidth]{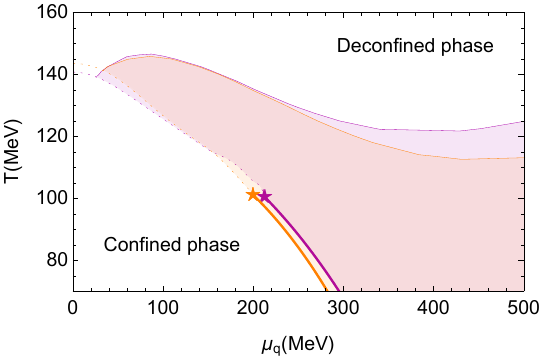}
\\
\includegraphics[height=0.042\textwidth]{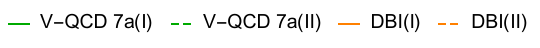}
\\
\includegraphics[height=0.32\textwidth]{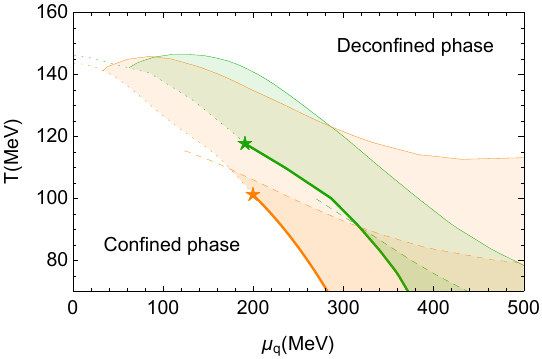}
\caption{\small Phase diagram with shaded areas indicating instability regions, where thick solid curves, dotted curves, and stars represent first-order phase transitions, crossovers, and CEPs, respectively. Dashed curves in the bottom plot illustrate the flavor-dependent suppression of the upper boundary of these regions. From top to bottom: the panels compare different V-QCD potentials, DBI versus EYMD models, and the influence of flavor dependence on instability suppression.}
    \label{pcomb}
    \label{fig:pcomb}
\end{figure}

We then construct numerically the charged black hole backgrounds at relevant temperatures and chemical potentials, and check whether the quasi-normal mode solutions to~\eqref{eq:LRinstabilityeq} show an instability of the type shown in Fig.~\ref{fig:p0}. The results for the extent of the instability are shown in Fig.~\ref{fig:pcomb} as the shaded areas. They are li\-mi\-ted from below by the first-order phase transition (thick solid curves) or crossover (dotted curves) to the confined phase. The transition curves are estimated as detailed in~\cite{Demircik:2021zll} for the V-QCD models and in~\cite{JJP} for the other models, \ie, the DBI and its leading order truncation, the Einstein--Yang--Mills--dilaton (EYMD) models.

The striking observation is that the unstable region extends to high temperatures, close to 150~MeV, and to low chemical potentials, entering the region of $\muq/T \lesssim 1$ where lattice data is available~\cite{Philipsen:2005mj,deForcrand:2009zkb,Guenther:2017hnx,Bazavov:2017dus,Ding:2015ona}. In particular, the instability is found before reaching the CEPs marked as stars. The length scale of the instability is estimated~\cite{QNMpaper} to be around $0.35$~fm in this region. Given that the instability extends to surprisingly low densities, it is crucial to analyze in detail what are possible sources of uncertainty in our computations. 

A rather obvious caveat in our approach is that we are only analyzing the instability in the black hole phases of the dual geometry, which is expected to be the correct description for the deconfined phase~\cite{Witten:1998zw,Aharony:1999ti}.  Therefore results may not be quantitatively reliable as we approach the crossover region, and we do not even attempt to predict the extent of the instability below the crossover curves. Note however that we also observe a strong instability in parts of the phase diagram that appear to be purely deconfining, such as the regions above the first order phase transition and the CEP. 

The uncertainty due to the fitting procedure can be estimated by comparing the results for the three different parametrizations of the V-QCD model (top panel of Fig.~\ref{fig:pcomb}) and the results in the DBI model, where a different fitting strategy was used (shown in orange in the middle panel and compared to the 7a fit of the V-QCD model in the bottom panel of Fig.~\ref{fig:pcomb}). We see that while there are small differences between the models, the qualitative behavior is unchanged by a sizable margin in the low-density region (say, for $\muq \lesssim 250$~MeV). Note that we have zoomed into the most relevant region in the plots which enhances the differences. The uncertainties associated with fitting are approximately an order of magnitude too small to impact our conclusions, particularly in the low-density regime.

Another choice of uncertainty is the choice of action, in particular the flavor term in~\eqref{sf}. This can be addressed by comparing the results between the DBI model and its leading-order truncation, the EYMD model, shown in the middle panel of Fig.~\ref{pcomb}. The model dependence is even weaker than the fit dependence discussed above. This is in agreement with 
the results for the phase diagram and the EoS in the homogeneous phase, which turn out to be almost independent of the choice of the flavor term, as indicated by the close agreement of the phase transition lines and the CEP between the two models in this plot (see also~\cite{JJP}).

We find, however, that the most important source of uncertainty in our computation is the dependence on flavors. In order to understand this, we analyze the structure of the fluctuation equation~\eqref{eq:LRinstabilityeq}.
The CS term introduces fluctuations through the final term in equation (\ref{eq:LRinstabilityeq}), acting as a negative mass term. This contribution can lead the system towards instability when its strength reaches a sufficient threshold.
There are two factors enhancing the strength of this term. First, it is proportional to the quark number density, $\hat{n}$. Second, it is inversely proportional to the fourth power of the potential $w(\phi)$. 
This potential encodes the finite-density effects and it is fitted to the baryon number susceptibility $\chiB$.  
The susceptibility decreases significantly with decreasing temperature \cite{Borsanyi:2011sw}. This implies that $w(\phi)$ decreases rapidly with 
$\phi$. The largest $\phi$ values are reached by the black hole solutions with small chemical potential and temperature values in the CEP region \cite{Hoyos:2021njg,Alho:2012mh,Alho:2013hsa}, which enhance the strength of CS term in the same region via $w(\phi)^4$ in the denominator. When the combination of these two enhancing terms is taken into account, the shape of the instability region can be understood.

The quark number susceptibilities have been computed for different flavors on the lattice in~\cite{Borsanyi:2011sw,HotQCD:2012fhj}, with the result that light quark susceptibilities are significantly higher than the strange quark susceptibility in the crossover region. This suggests that $w$ and therefore also the instability depends strongly on the flavor. However, it is not straightforward to estimate this effect in our model, because the lattice data assumes $\Nf=2+1$ flavors, \ie, a massive strange quark, and including it in holographic models is challenging. But in order to obtain an estimate of the size of the effect, we adjust the $w$ functions of the V-QCD 7a and DBI models such that they fit the light quark susceptibility of~\cite{Borsanyi:2011sw} rather than $\chiB$ (see Supplemental Material).

The result of this estimate is shown in the bottom plot of Fig.~\ref{pcomb}. The onset of the instability is marked by thin solid (dashed) curves for the original fits to quark number susceptibility (the fits to light quark susceptibility), respectively.  The results demonstrate a suppression in the instability region  when fitted to the light quark susceptibility: the onset temperatures are lowered by roughly 20\%. This is enough to remove the instability near our estimate for the CEP for the V-QCD model. However we remark that i) our estimate is naive, since the effect due to the strange quark mass is not isolated, ii) the effect of the flavor dependence to the CEP was not taken into account, and iii) the data for the strange quark susceptibility is lower than the total quark number susceptibility~\cite{Borsanyi:2011sw}, indicating an opposite effect, \ie, enhancement of the instability in the strange quark sector.

\section{Discussion} 

We demonstrated that a modulated instability, which appears at low density near the critical crossover temperature of QCD and most likely cloaks the CEP in the phase diagram, is unavoidable in holographic models fitted to the unflavored lattice QCD data. Apart from the various models considered in this letter, this applies at least to the models of~\cite{Knaute:2017opk,Critelli:2017oub,Li:2023mpv} due to the mild dependence on the details of the fit. Such models have been widely used to study the CEP of QCD~\cite{Hippert:2023bel,Fu:2024wkn} and the EoS in general~\cite{Rougemont:2023gfz}. 
We identified the flavor dependence, and in particular effects due to the strange quark mass, as the main uncertainty in our analysis. It is therefore critical to extend our analysis to take into account such effects.

Given the weak model dependence, our study strongly suggests that this instability is not limited to holographic modeling, but QCD at physical quark masses shows a corresponding inhomogeneous phase in this region, characterized by nonzero and spatially dependent chiral currents. The observed instability roughly coincides with the area where lattice studies detect a breakdown in the Taylor expansions (see, \eg, \cite{Gavai:2008zr,Karsch:2010hm,Borsanyi:2021sxv,Mondal:2021jxk}), hinting at a possible link that merits further investigation. The modulated ground state could potentially be constructed on the lattice or observed in experiments such as Phase II of BES, given that the timescale $\tau=\text{Im}\omega_\mt{max}$ for the mode $\sim e^{t/\tau}$ driving the instability is notably short, approximately $0.2$ fm/c. Additionally, the inhomogeneous phase likely exhibits unique transport properties, opening new avenues for experimental and theoretical investigation.

\emph{Acknowledgements.}---%
We thank J.~Cruz Rojas, U.~G\"ursoy, C.~Hoyos, T.~Ishii, K.~Kajantie, E.~Kiritsis, T.~Mitra, S.~Nakamura, F.~Nitti, F.~Pe\~na-Ben\'\i{}tez,  E.~Pr\'eau, M.~Roberts, K.~Rummukainen, D.~Schaich,  A.~Schmitt, and A.~Vuorinen for useful discussions. T.~D. acknowledges the support of the Narodowe Centrum Nauki (NCN) Sonata Bis Grant no. 2019/34/E/ST3/00405. N.~J. has been supported in part by Research Council of Finland grants no. 345070 and 354533. M.~J. has been supported by an appointment to the JRG Program at the APCTP through the Science and Technology Promotion Fund and Lottery Fund of the Korean Government and by the Korean Local Governments -- Gyeong\-sang\-buk-do Province and Pohang City -- and by the National Research Foundation of Korea (NRF) funded by the Korean government (MSIT) (grant no. 2021R1A2C1010834). A.~P. acknowledges the support from the Jenny and Antti Wihuri Foundation.

\bibliography{refs.bib}

\onecolumngrid
\newpage
\appendix

\centerline{\bf \Large Supplemental Material}

\vspace{7mm}

Here we discuss the details of the fitting of the $w(\phi)$ function, \ie, the coupling of the gauge fields in the action, and how we estimate its dependence on the mass of the strange quark. As we explained in the main text, the choice of this function is critical for the extent of the instability.

The form of the $w(\phi)$ is determined by fitting to the susceptibility lattice data, \eg, the (dimensionless) baryon number susceptibility
\be
 \chiB = \frac{1}{\Nc^2T^2} \frac{\partial^2p}{\partial \muq^2}\ ,
\ee
where the temperature and other chemical potentials are held fixed in the partial derivative. In the flavor-independent setup, this can be directly compared to the expression obtained from the holographic model:
\be \label{eq:chiBholo}
 \chiB = M_\mathrm{p}^3  \left[\int_0^{r_\mathrm{h}} \mathrm{d}r \ \frac{1}{e^{A(r)}\Vf(\phi(r))w(\phi(r))^2} \right]^{-1} \ .
\ee
Here $r_\mathrm{h}$ denotes the location of the horizon and the boundary, where the field theory lives, located at $r=0$. The form of $\Vf$ is determined by a fit to lattice data for the equation of state, so comparison to the data for $\chiB$ only determines the shape of $w(\phi)$~\cite{Jokela:2018ers,JJP}. We stress that it is the same combination of functions, \ie, $\Vf(\phi)w(\phi)^2$, which appears in the expression~\eqref{eq:chiBholo} and the denominator of the Chern--Simons contribution in~\eqref{eq:LRinstabilityeq}, so that the strength of the instability is directly linked to the fit of the susceptibility.

\begin{figure} [h!] 
\includegraphics[width=0.7\textwidth]{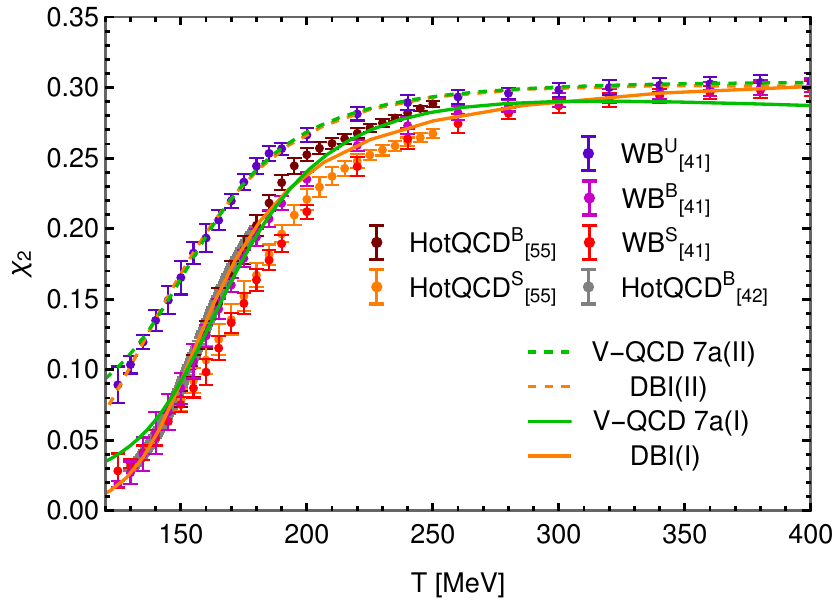}
\caption{\small We estimate the impact of the strange quark mass on the $w(\phi)$ fits. The dashed curves correspond to fitting the potentials to only the light quark susceptibility. The solid lines correspond to fitting the potentials to the $\chiB$. All susceptibilities, light quark (U), baryon (B), and strangeness (S), of \cite{Borsanyi:2011sw} are shown. Notice that the light quark and strangeness susceptibilities are divided by three. The brown and orange data points are the baryon and strangeness susceptibilities from~\cite{HotQCD:2012fhj}, while the gray data is $\chiB$ of \cite{Bazavov:2020bjn}. 
}
    \label{fig:suscis}
\end{figure}

However, as pointed out in the text, the lattice data~\cite{Borsanyi:2011sw,HotQCD:2012fhj} shows significant dependence on the quark flavors. In Fig.~\ref{fig:suscis}, we reproduce the relevant lattice data, adjusting the normalizations of different susceptibilities such that they can be directly compared. While the flavor dependence, which mostly arises due to the strange quark mass, is relatively weak, it is important in our case because of the sensitivity of the instability to the susceptibility fit. We cannot do a consistent fit to all data because the dependence of the holographic models on the strange quark mass has not yet been consistently included in the holographic models we are using.

What we can do here is to estimate the size of the flavor effects by carrying out, in addition to the standard fit to the baryon number susceptibility, a fit to the light quark susceptibility (the violet points and error bars in Fig.~\ref{fig:suscis}). This gives an estimate on the sensitivity of our results to the susceptibility fit in the light quark sector. We  present here the fit results to both baryon number and light quark susceptibilities for the 7a variant of the V-QCD model and the DBI model in the second approach discussed in the main text, where there is no phase transition and the data is fitted at all temperatures.

The functional form  of $w(\phi)$ in V-QCD model is given as~\cite{Jokela:2018ers}
\begin{align} \label{w1}
 \frac{1}{w(\phi)} &=
w_0\left[1+\frac{w_{\text{U}}s_1 e^\phi/\lambda_0}{1+s_1 e^\phi/\lambda_0}+w_{\text{IR}}
e^{-\lambda_0/(s_2 e^\phi)}\frac{(s_2 e^\phi/\lambda_0)^{4/3}}{\log(1+s_2 e^\phi /\lambda_0)}
 \right]\ , \quad \lambda_0=8\pi^2\ ,
 \end{align}
and $w(\phi)$ in DBI model is~\cite{JJP}
\begin{align} \label{w2}
 \frac{1}{w(\phi)} &=
w_0 \cosh(\gamma_2 \phi/\phi_0)+w_2  (\phi/\phi_0)^2+w_4 (\phi/\phi_0)^4+w_6 (\phi/\phi_0)^6 \ .  
\end{align}
Here the parameter $\phi_0 = \sqrt{3/8}$ was added due to the change of normalization conventions with respect to~\cite{JJP}.
The parameter sets for V-QCD potential 7a and DBI model are given in Table~\ref{parasets} for the fits with baryon (or quark) number susceptibility (I) and light quark susceptibility (II). In Fig.~\ref{fig:suscis}, the comparison of the corresponding fits with the lattice data \cite{Borsanyi:2011sw,Bazavov:2020bjn} is presented.

\begin{table}[h!]
\centering
\begin{tabular}{c c}
\begin{tabular}{|c | c | c | c | c | c |} 
 \hline
  &  $w_0$  & $w_\text{U}$  & $w_\text{IR}$  & $s_1$ & $s_2$  \\ 
 \hline  \hline
 7a(I)  & 1.280 & 0 & 18  & 3 & 1.18 \\
 \hline
 7a(II)  & 1.015 & 0.9 & 10 & 3 & 1.2 \\
 \hline
\end{tabular}
\hspace{10mm}%
\begin{tabular}{|c | c | c | c | c | c |} 
 \hline
  &  $w_0$  & $\gamma_2$  & $w_2$ & $w_4$ & $w_6$   \\ 
 \hline  \hline
 DBI(I)  & 3.009 & 1.087 & -1.792 & -0.166 & -0.00743  \\
 \hline
 DBI(II)  & 3.277 & 0.954 & -1.688 & -0.0770 & -0.00508  \\
 \hline
\end{tabular}
\end{tabular}
  \caption{The parameter sets of $w(\phi)$ potentials for 7a(I), 7a(II), DBI(I), and DBI(II).}
    \label{parasets}
\end{table}

\end{document}